\DeclareRobustCommand{\baselinestretch{2}}

\documentclass[12pt,letterpaper]{article}                  
\usepackage{graphics}

\begin{document}

\title{Distortion-free tight confinement and step-like decay of fs pulses in free space}

\author{G. Nyitray and S. V. Kukhlevsky }

\date{\textit{Department of Experimental Physics, Institute of Physics, \\
  University of Pecs, Ifjusag u. 6, H-7624 Pecs, Hungary}}

\maketitle 

\begin{abstract}A method of formation of the tightly confined distortion-free fs pulses with the
step-like decreasing of intensity under the finite-length
propagation in free space is described. Such pulses are formed by
the Fresnel source of a high refraction-index waveguide. The
source reproduces in free space a propagation-invariant
(distortion-free) pulse confined by the waveguide. Converse to the
case of material waveguides, when the pulse goes out from the
Fresnel (virtual) waveguide its shape is not changed, but the
intensity immediately drops down to the near-zero level.
\newline
OCIS numbers: 320.0320, 320.2250, 320.5540, 320.5550.
\end{abstract}


Tight transverse confinement of a light wave in free space
together with fast decreasing of its intensity after the
finite-length propagation are required in many fields of optics,
such as non-destroying light-matter interaction, control of light
penetration-depth and surface processing. The effect is usually
achieved by strong focusing a light beam by short-curvature-radius
lenses or spherical mirrors. In the case of fs pulses, the control
of the transverse and longitudinal intensity distributions is a
particularly difficult problem, because the ultrashort pulses are
distorted in the space and time domains by such optical elements
(see, for example Refs. \cite{Kemp,Bor,Jack,Bert,Niso,Ferm}).
Recently, it was shown that the tightly confined
propagation-invariant (distortion-free) continuous waves with the
fast decreasing of the intensity after the finite-length
propagation in free space could be formed by the Fresnel source of
a high refractive-index hollow waveguide having total-reflection
walls \cite{Kuk1,Cann}. If such an effect exists for the
ultrashort pulses, this could be very important for the fs optics
and applications. In the present article, the formation of tightly
confined distortion-free fs pulses with the step-like decreasing
of intensity under finite-length propagation in free space is
described.

Let us describe the Fresnel source that reproduces in free space a
propagation-invariant (distortion-free) pulse tightly confined by
a high refractive-index hollow waveguide with total-reflection
walls. The plane-parallel guide having the length $L_w$ and width
$2a$ is considered. A propagation-invariant time-harmonic wave
$E'(x',z,t,{\omega})$ is supported in the free-space by the
constructive interference of multiple beams
$E'_m(x',z,t,{\omega})$ launching from the Fresnel zones
$[x_{m}^{min},x_m^{max}]$ of the virtual source of the guide
\cite{Kuk2}:
\begin{eqnarray}
E'(x',z,t,{\omega})=\sum_{m=-M}^{M}E'_m(x',z,t,{\omega}),
\end{eqnarray}
with
\begin{eqnarray}
E'_m(x',z,t,{\omega})={\frac{1}{\sqrt{2{i}\lambda}}}\int_{x_m^{min}}^{x_m^{max}}
{\frac{\exp{[ik{r(x',x)}]}}{r(x',x)}}\nonumber \\
{(1+\cos\Theta_m){E_{m}(x,0,t,{\omega})}dx},
\end{eqnarray}
where $2M+1$ is the number of zones (beams) of the Fresnel source
that contribute the energy into the field $E'(x',z,t,{\omega})$;
${\mid}x_m^{max}{\mid}=(2m+1)a$ and
${\mid}x_m^{min}{\mid}=(2m-1)a$; ${\omega}=kc$ is the wave
frequency; $r(x',x)$ is the distance between points $x'$ and $x$;
$E_m(x,0,t,{\omega})=E_0([x-2am](-1)^m,0,t,{\omega})$, where
$E_0(x,0,t,{\omega})$ is the field at the guide entrance $(z=0)$.
The value $M=M(z,d_M)$ is determined by the transverse dimension
$d_M(k,a,z)$ of the beam $E'_M(x',z,t,{\omega})$ at the guide exit
$(z=L_w)$. By analogy with the Fresnel lens, the virtual source of
the waveguide is called the Fresnel waveguide \cite{Cann}. The
source reproduces in free space a diffraction-free beam
\cite{Brit,Ziol,Durn,Gori,Lu} confined by the waveguide. In the
case of an ultrashort pulse $E'(x',z,t)$ guided by the waveguide,
the field of the pulse at the guide entrance is represented in the
form of a Fourier integral $F[E(x,z=0,{\omega})]$. Using the
Fresnel-waveguide representation for the Fourier components and
substituting the result into the Fourier integral we get the
propagation-invariant pulse $E'(x',z,t)$ in free space
\cite{Kuk2}. The pulse is supported in free space by the
constructive interference of the $2M(z)+1$ ultrashort-pulses
launching from the Fresnel source of the waveguide.

The Fresnel source reproduces in free space a distortion-free
(propagation-invariant) pulse confined by the waveguide. Converse
to the case of material waveguides \cite{Marc,Oka}, when the pulse
goes out from the Fresnel (virtual) waveguide its shape is not
changed, but the intensity immediately drops down to the near-zero
level. The effect, which looks like the instant decay or
annihilation of a tightly confined pulse, is illustrated by the
numerical examples in Figs. 1(a) and 1(b). Figure 1(a) shows the
intensity distribution ${\mid}E(x,z){\mid}^2$ on the axis $(x=0)$
of the virtual waveguide computed for the pulse having the two
different durations $\tau$: 100 and 150 fs. The intensity
distribution of the continuous wave $({\tau}=\infty)$ is shown in
the figure for the comparison. Figure 1(b) demonstrates the
intensity distributions of the pulse (${\tau}$=150 fs) at the
points B, C and D. We notice that the pulse propagates in the
region of the virtual waveguide as the superposition of many
pulses that diffract in the off-axis direction and interfere with
each other. Analysis of the intensity distributions indicates
existence of the two main stages of the evolution of the pulse
under the free-space propagation: the propagation-invariant AC
$(z<L_w)$ and collapse CD $(z>L_w)$ stages. In the region AC
$(z<L_w)$, the intensity distribution of the pulse is practically
unchanged that demonstrates the propagation-invariant properties
of the pulse. The pulse amplitude is constant in the region AB and
shows some oscillations in the part BC of the region AC. The
amplitude of these oscillations, which increases with increasing
the pulse duration, reaches maximum value before the pulse decay.
In the collapse region CD $(z>L_w)$, the number of pulses
$2M(z=Lw)+1$ is not sufficient for supporting of the
propagation-invariant field distribution. In this region, the
pulse shape is not changed, but the intensity immediately drops
down to the near-zero level. The pulse decay is caused by the
destructive interference between the pulses launched from the
finite-width Fresnel source of the finite-length waveguide. The
distance $AC=L_w$ strongly correlates with the width
$(2M(z=L_w)+1)2a$ of the Fresnel source. This presents a method of
formation of the tightly confined distortion-free fs pulses with
the step-like decreasing of intensity under the finite-length
propagation in free space. The propagation length AC could be
controlled by variation of the width of the Fresnel source. The
above-described theoretical principle of the Fresnel source can be
realized experimentally by using the $2M(z)+1$ fs-pulses. In the
simplest case, the source can be formed by the interference of
only two fs-pulses propagating in the off-axis direction, as shown
in Fig. 2. Notice, that the two-pulse technique is similar to the
method of formation of the dynamical gratings (see, for example
Refs. \cite{Mazn,Goodno}).

In conclusion, a method of formation of the tightly confined
distortion-free fs pulses with the step-like decreasing of
intensity under the finite-length propagation in free space was
described. The pulses are formed by the Fresnel source of a high
refraction-index waveguide. The length of the distortion-free
propagation and the time of the pulse decay are controlled by
variation of the width of the Fresnel-waveguide source. It should
be noted that the method can be extended to the guiding of fs
pulses in free-space by the 3-dimensional virtual waveguides. In
this case, one could use the Fresnel sources of 3-dimensional
high-refraction-index waveguides \cite{Kuk3}.
%

This study was supported by the Fifth Framework of the European
Commission (Financial support from the EC for shared-cost RTD
actions: research and technological development projects,
demonstration projects and combined projects. Contract
NG6RD-CT-2001-00602). The authors thank the Computing Services
Center, Faculty of Science, University of Pecs, for providing
computational resources.
%

\newpage

\section*{List of Figure Captions}

Fig. 1. (a) The step-like evolution of the normalized intensity of
the Gaussian-shaped ultrashort light pulse on the
virtual-waveguide axis under the free-space propagation computed
for the two different pulse durations $\tau$ :  (1) - 100 fs and
(2) - 150 fs. The intensity distribution of the continuous wave
(3) and the step-like function (step(z)) are shown for the
comparison. (b) The normalized intensity distributions of the
propagation invariant (distortion-free) pulse ($\tau$ = 150 fs) at
the points B, C and D. The white dashed line indicates the region
of the virtual waveguide. The guided pulse is supported in the
free space by the diffraction and interference of $2M+1$ pulses of
the Fresnel-waveguide source. The intensity distributions were
calculated using the following parameters: ${\lambda}_0$ = 500 nm,
$z{\in}$[0.05 m, 4.05 m], $x{\in}$[-750 ${\mu}$m, 750 ${\mu}$m],
${\tau}{\in}$[-750 fs, 750 fs] and $M=71$, where ${\lambda}_0$ is
the he central wavelength of the pulse (wave-packet).

\noindent Fig. 2. The schematic diagram of the Fresnel-wavegiude
source formed by the interference of the two fs-pulses ($p_1$ and
$p_2$) propagating in the off-axis direction. The box indicated by
the white dashed line shows the region of the virtual waveguide.

\newpage

\vspace{3cm}

  \begin{figure}[h]\centerline{\scalebox{0.6}{\includegraphics{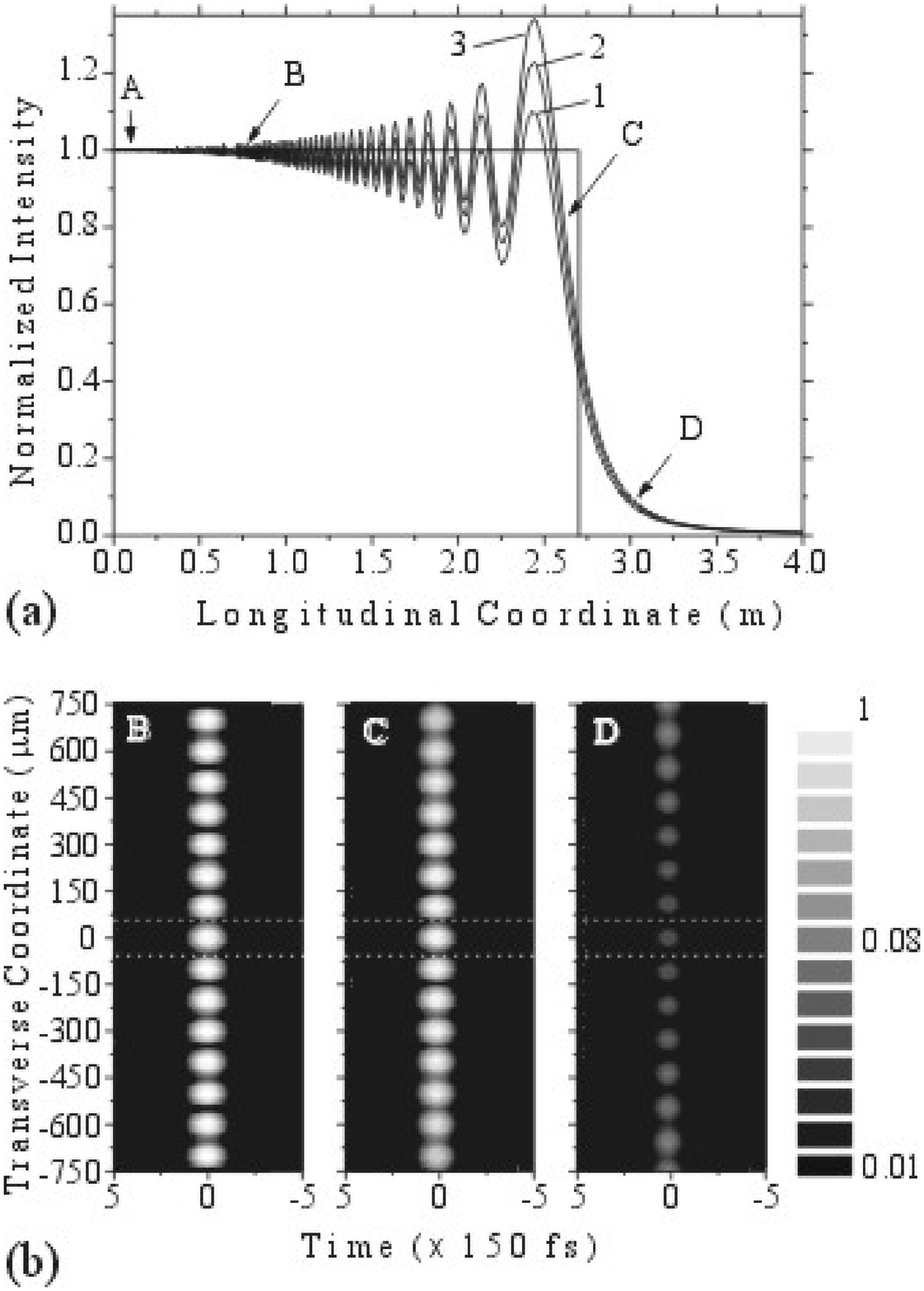}}}
  \caption{}
  \end{figure}

\newpage

\vspace{3cm}

  \begin{figure}[h]\centerline{\scalebox{0.7}{\includegraphics{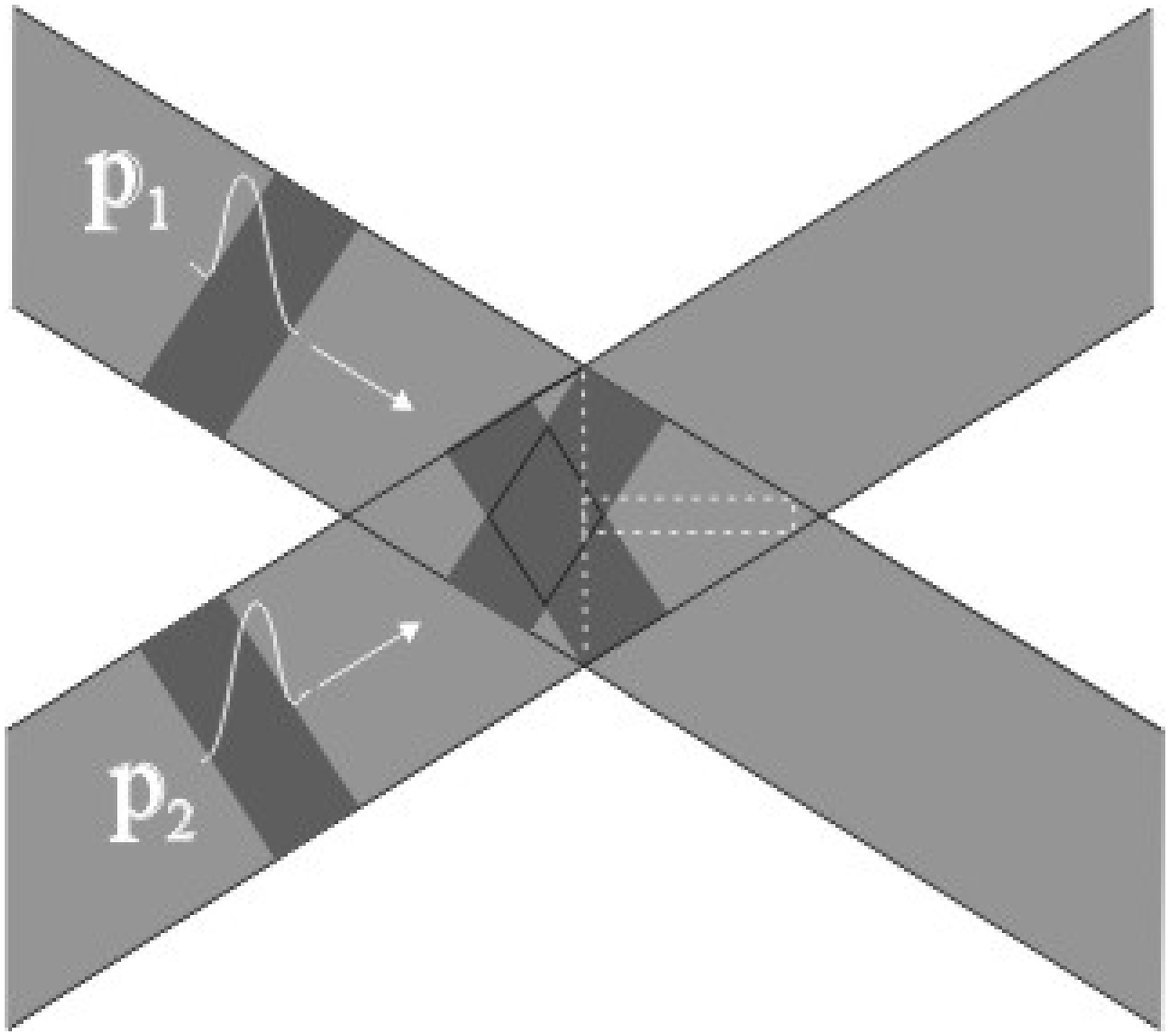}}}
  \caption{}
  \end{figure}

\end{document}